\documentclass[%
 aip,
 amsmath,amssymb,
 reprint,%
]{revtex4-1}

\usepackage{graphicx}% Include figure files
\usepackage{dcolumn}% Align table columns on decimal point
\usepackage{bm}% bold math
\usepackage[utf8]{inputenc}
\usepackage[T1]{fontenc}
\usepackage{mathptmx}

\usepackage{notes2bib}

\begin{document}

\preprint{AIP/123-QED}

\title{Atomic Layer Deposition of SiO$_{2}$-GeO$_{2}$ multilayers}

\author{Jordi Antoja-Lleonart}
 \email{j.antoja.lleonart@rug.nl}
\author{Silang Zhou}
\affiliation{Zernike Institute for Advnaced Materials, University of Groningen, 9747AG Groningen, The Netherlands}

\author{Kit de Hond}
\author{Gertjan Koster}
\author{Guus Rijnders}
\affiliation{MESA+ Institute for Nanotechnology, University of Twente, PO Box 217, 7522 NH Enschede, The Netherlands}

\author{Beatriz Noheda}
 \email{b.noheda@rug.nl}
\affiliation{Zernike Institute for Advanced Materials, University of Groningen, 9747AG Groningen, The Netherlands}

\date{\today}

\begin{abstract}
Despite its interest for CMOS applications, Atomic Layer Deposition (ALD) of GeO$_{2}$ thin films, by itself or in combination with SiO$_{2}$, has not been widely investigated yet. Here we report the ALD growth of SiO$_{2}$/GeO$_{2}$ multilayers on Silicon substrates using a so far unreported Ge precursor. The characterization of multilayers with various periodicities reveals successful layer-by-layer growth with electron density contrast and absence of chemical intermixing, down to a periodicity of 2 atomic layers.
\end{abstract}

\maketitle

In the last four decades, Atomic Layer Deposition (ALD) has seen widespread adoption as a thin film growth technique \cite{Puurunen2005,Johnson2014}. Its scalability, unprecedented conformality, and thickness control down to the atomic level, all make it a valuable asset to most nanofabrication efforts, playing a key role in commercial semiconductor manufacturing. Though chiefly known for the growth of relatively simple compounds, such as binary oxides, nitrides or sulfides, amorphous for the most part, ALD is also used to grow metals \cite{Hamalainen2014}, and recently more complex materials \cite{Coll2019}, including perovskites \cite{Tyunina2008, Sbrockey2012, Akbashev2014, McDaniel2015, Lin2019} have been realized as well. 

One of the materials with the longest history using the ALD techniques is SiO$_{2}$\bibnote{Occasionally "pulsed metal organic chemical vapor deposition" is used instead of ALD, though in practice both methods are very similar.}, a key element in the microelectronics industry with applications as passivation layer and gate oxide, among others. Less ubiquitous is the ALD growth of the related material GeO$_{2}$; its growth using ALD is relatively unexplored and not many of its possible precursors have been tested \cite{Chalker2012,Adinolfi2019,Matero2014}. Research on GeO$_{2}$ films has been mainly devoted to the study of the GeO$_{2}$/Ge interface, with GeO$_{2}$ films being proposed as a means to reduce the concentration of interface states between Ge and a high-K dielectric on top \cite{Yoshida2014, Shibayama2015, Kanematsu2016}, with the goal of realizing MOSFETs with a Ge-based channel. In these studies, thermal or plasma oxidation, as well as vapor growth \cite{Matsubara2008, Robertson2015}, were used. It is worth mentioning that these works use precursors containing alkoxy or halide ligands, which give rise to comparatively slow reaction rates.

Thin films consisting of SiO$_{2}$ and GeO$_{2}$ multilayers have been investigated in the past, both from solution and vapor deposition methods\cite{Zhang2003,Ho2004,Ho2005}, with the focus being mostly on their optical properties. In this work, we show that using tetrakis(dimethylamino) germanium (IV) (TDMAGe) as precursor, it is possible to deposit GeO$_{2}$, as well as SiO$_{2}$/GeO$_{2}$ multilayers by thermal ALD.

We use a Picosun R-200 Advanced hot-wall ALD System whose chamber opens to a glovebox containing a nitrogen atmosphere, with controlled oxygen and water concentrations. We grow oxide thin films using organometallic Si and Ge precursors. Respectively, these are bis(diethylamino) silane (BDEAS, commonly known as SAM-24) and tetrakis(dimethylamino) germanium (IV) (TDMAGe), both of them purchased from Air Liquide. The precursors are housed in canisters from Picosun, PicohotTM 300 and PicohotTM 200, which allow heating up to 260ºC and 200ºC, respectively. The organometallic precursors are delivered into the reaction chamber, through heated valve blocks, using nitrogen as carrier gas.

The oxidizer used in this work is ozone, which is produced from an INUSA Ozone Generator using Oxygen 6.0. Our valves allow a minimum opening time of 0.1s. Accessible process temperatures range from 100ºC to 300ºC, and the typical process pressure is 17 hPa. A temperature of 38ºC for the BDEAS precursor bottle was established to give acceptable delivery rates. The TDMAGe bottle needed to be heated to 80ºC to achieve similar precursor delivery rates to the reaction chamber. The gas lines downstreaming from the bottles were heated to 10-20ºC above the temperature of their respective bottle to avoid precursor condensation taking place before reaching the reaction chamber.

The SiO$_{2}$ growth from BDEAS was optimized in collaboration with Picosun{\textregistered}.The reactor temperature was set to 300ºC. The BDEAS pulse length was 0.1 s, followed by a 6.0 s N$_{2}$ purge. The ozone pulse length was 8.0 s, also followed by a 6.0 s N$_{2}$ purge. In our system, SiO$_2$ grown in this way shows a growth per cycle (GPC) of about 0.7Å. The GeO$_{2}$ growth has been independently optimized in the present work, as detailed below. The films are grown on 15x15 mm$^2$ square pieces of Si(100) wafers purchased from Microchemicals GmbH. 

Thickness determination has been performed by elllipsometry using a J.A. Woollam Co. V-VASE system. In order to validate the ellipsometry results and to investigate the quality of the multilayers, X-ray reflectivity (XRR) has also been performed using a Bruker D8 Discover diffractometer with a high brilliance Cu rotating anode generator. The topographic features of the samples and their roughness were imaged by a Bruker Dimension Icon Atomic Force Microscope (AFM). Details on these methods are included in the Supplementary Material.

Although the growth of GeO$_{2}$ by ALD using TDMAGe as a precursor was recently patented by ASM \cite{Matero2014}, the details of the growth were not reported. Even though the precursors used for GeO$_{2}$ and SiO$_{2}$ are quite different, both of them use alkylamine ligands. This allowed us to optimize the GeO$_2$ growth, using as starting parameters those of the  SiO$_{2}$ growth.

\begin{figure}
\includegraphics[width=0.47\textwidth]{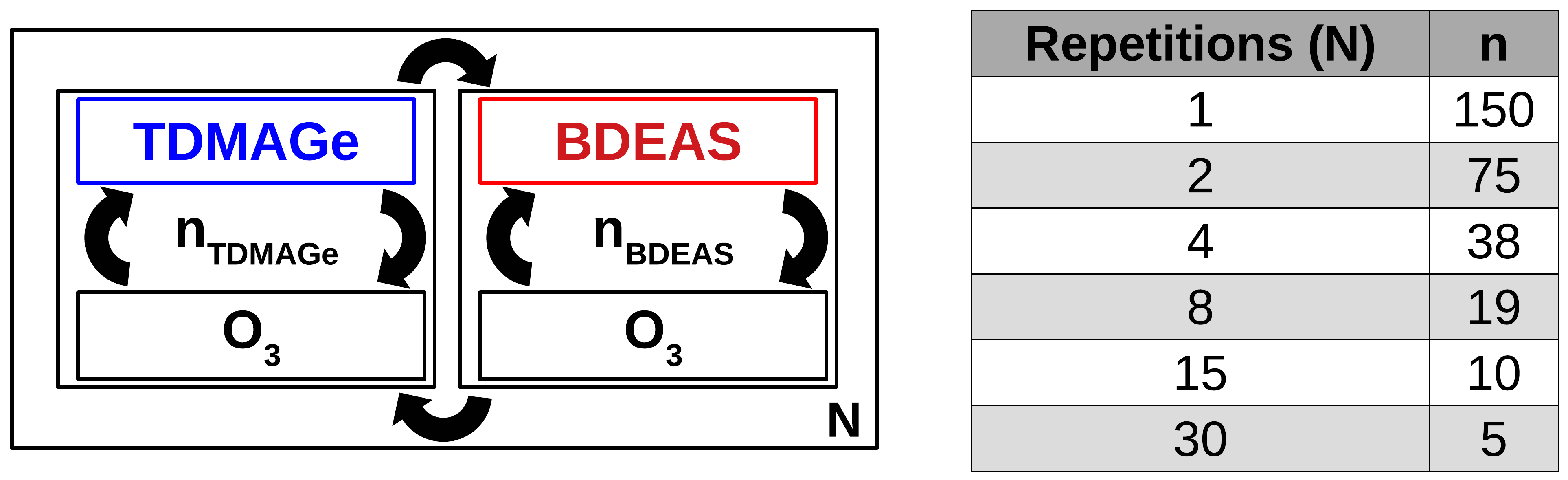}
\includegraphics[width=0.47\textwidth]{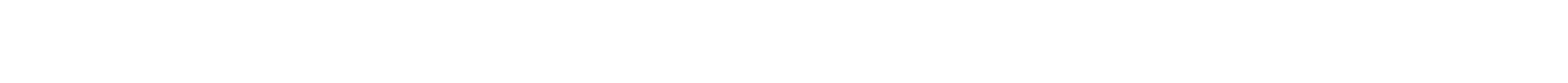}
\includegraphics[width=0.47\textwidth]{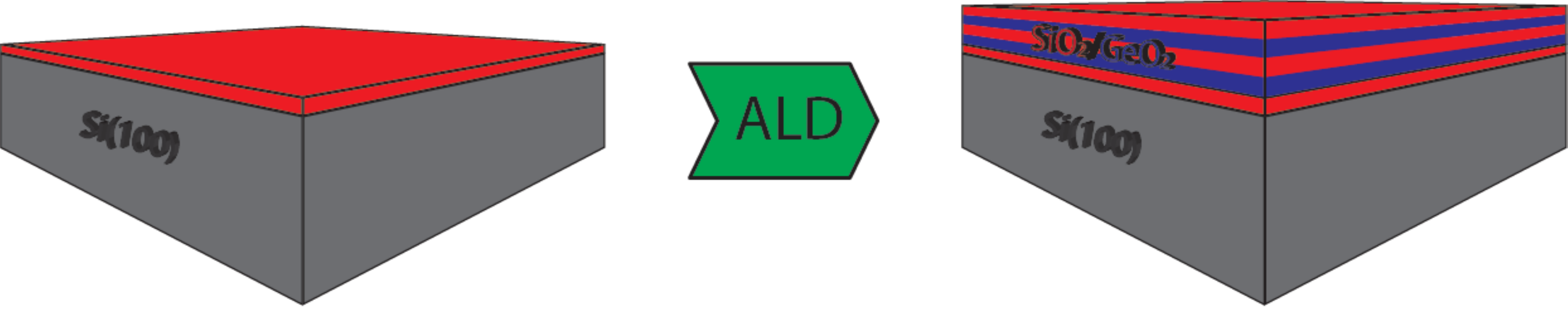}
\caption{\label{fig:Crystallizationsketch}Top left: Schematic representation of the synthesis method for the mixed oxide films. In our study, the pulse trains for the two precursors contain the same number of pulses: n$_{TDMAGe}$=  n$_{BDEAS}$= n, and it varies from film to film. Top right: Table detailing the number of precursor pulses in each train (n) per sublayer and the total number of cycles (N) in the different films. For the 1:1 pulse ratio used in this work, the total number of cycles (2nxN) was kept to approximately 300 for the whole series. Bottom: Sketch of the expected layered structure after growth.}
\end{figure}

In the case of the combined SiO$_{2}$/GeO$_{2}$ multilayer growth, there are practical constraints to the process. First, the well-known SiO$_{2}$ precursor BDEAS\cite{Won2010,ONeill2011,Baek2012,Mallikarjunan2015,Fang2016,Schwille2017}, requires ozone or oxygen to function properly in thermal ALD. If water vapor is used instead, the Si-H bonds in the precursor do not react, which results in decreased growth rate and, possibly, leading to too high impurity concentrations in the film. For this reason, it is highly desirable to simplify the process by using ozone as the oxidizer in GeO$_{2}$ growth as well, even though this may lead to combustion-like reactions and less gentle oxidization. The ozone pulse length was fixed at 8s, sufficient to ensure a complete half-reaction.

Second, the SiO$_{2}$ growth is optimal at or above 300ºC. When growing subsequent layers of different materials by ALD, it is in principle possible to change the reaction temperature when switching from one oxide to the next. However cooling and heating the reactor, even for relatively small temperature differences, are slow processes, making the growth time prohibitively long if the temperature needs to be changed repeatedly. For this reason, while the optimal growth temperature for pure GeO$_{2}$ is determined, if GeO$_{2}$ growth is still acceptable at 300ºC, this temperature needs to be maintained in the multilayer growth.

\begin{figure}
\includegraphics[width=0.48\textwidth]{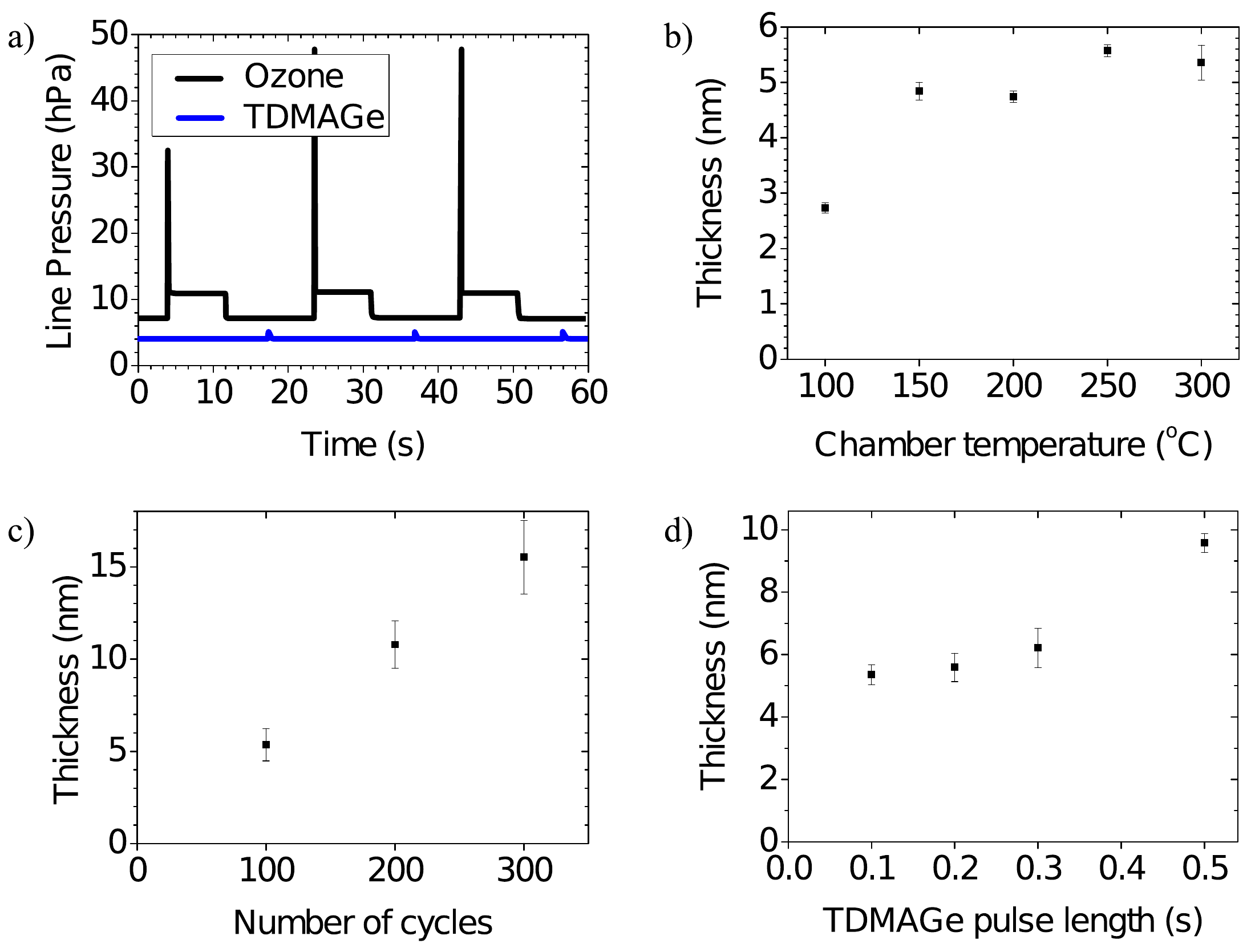}
\caption{\label{fig:ALDoptimize} a) Sketch of the pulse trains of the two precursors. b) GeO$_{2}$ film thickness after 100 cycles at various reactor temperatures. c) GeO$_{2}$ film thickness for different cycle numbers, at 300ºC. A linear regression for this regime yields a GPC of 0.51Å. d) GeO$_{2}$ film thickness for 100 cycles and at 300ºC for different TDMAGe pulse lengths. Error bars are estimates based on sample dispersion.}
\end{figure}

Our system minimum operating temperature is, for this process, about 100ºC. This is necessary to avoid condensation of the precursors or their reaction products in the chamber. Its maximum operating temperature is 300ºC in its current configuration. Within this range, the growth rate of pure GeO$_{2}$ increases with increasing temperature, remaining approximately invariant above 150ºC (Figure \ref{fig:ALDoptimize}). One explanation for this behaviour is that below 150ºC the chemisorption reaction rate for TDMAGe is too low for proper ALD behavior, resulting in a decreased GPC. It could be argued that the stabilization of the GPC up to 300ºC is an indication that precursor decomposition is still not significant at that temperature.

However, we can see that the GPC at 300ºC increases for longer TDMAGe pulse lengths (Figure \ref{fig:ALDoptimize}.d), while displaying approximately constant values for the short pulse lengths of 0.1s and 0.2s. This could indicate that this temperature is, in fact, sufficient to cause noticeable precursor decomposition, but that this has no noticeable effect in the growth provided that the TDMAGe pulses are short enough. This was confirmed by growing a film at 200ºC using TDMAGe pulses of 0.5s. In that case the GPC was 0.53Å. This shows that the effect of precursor decomposition on thickness can be minimized either by growing at sufficiently low temperatures, which is difficult in our case, as discussed before, or by keeping the pulse length short. Therefore, the shortest pulse length of 0.1s was chosen. We further show that the thickness of the grown films follows a linear dependence with the number of pulses (Figure \ref{fig:ALDoptimize}.c), as expected.

\begin{figure*}
\includegraphics[width=1\textwidth]{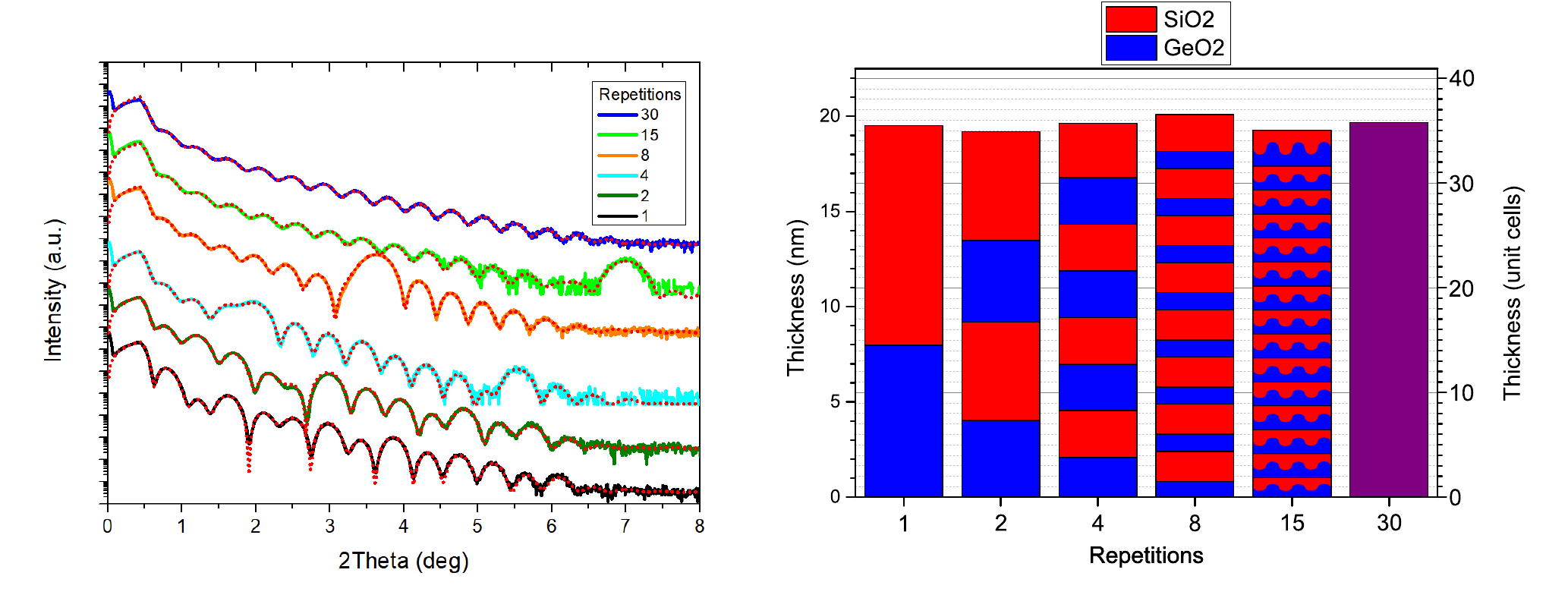}
\caption{\label{fig:SuperlatticeXRR} Left: XRR scans and fits (red dashed lines) of the six films, from which the thicknesses were extracted. Right: Side view of the grown films, showing the actual layer thicknesses as determined by XRR. The red and blue stripes indicate layers made of different oxides, whereas the purple stripe is used for a layer of mixed oxide. Superlattice fits add the constraint that all SiO$_{2}$ layers and all GeO$_{2}$ layers in a single heterostructure have respectively the same thickness, but this constraint is relaxed for the top SiO$_{2}$ and the bottom GeO$_{2}$ layer in each case. The native SiO$_{2}$ layers, though both measured by ellipsometry prior to the ALD process and accounted for in the models, are not depicted here. Note that the the models do include interface roughnesses. The right axis shows the thickness in multiples of 0.55nm, which is approximately the average between the SiO$_{2}$ and GeO$_{2}$ $\alpha$-quartz form c-parameter, an it is used as a characteristic length scale of the material (though it is amorphous in this case).}
\end{figure*}

We then proceeded to grow multilayer heterostructures with alternating SiO$_{2}$ and GeO$_{2}$ sublayers. This layered growth is often characterized in the literature by the ratio between the number of pulses of the metal precursors of the two components. Here we have synthesized multilayers with different SiO$_{2}$/GeO$_{2}$ periodicities using a constant 1:1 pulse ratio and the same total number of pulses (same total thickness), but varying the number of pulses in each train. In this way, it is possible to control the degree of intermixing during the growth step, without resorting to post-annealing.

For the experiments here, we set the total number of cycles to 300, of which half will be GeO$_{2}$ and the other half SiO$_{2}$. Note that, assuming that the behaviour displayed in Figure \ref{fig:ALDoptimize} can be extrapolated to lower cycle numbers, the expected thickness for the GeO$_{2}$ and SiO$_{2}$ layers would be 8 nm and 11 nm, respectively, taking into account our previously mentioned GPC values at the optimized process parameters. Thus, assuming ALD linear regime, we expect a total thickness of 19 nm approximately, without accounting for the native oxide present on the wafer.This is in good agreement with the values obtained from the fitting of the XRR patterns, which give total thicknesses ranging between 19.2 nm and 20.1 nm (see Table 1 in Supplementary Materials).

The first experiment, with 150 cycles of TDMAGe/O$_{3}$ followed by 150 cycles of BDEAS/O$_{3}$ is denoted as "one repetition" (see Figure \ref{fig:Crystallizationsketch}). One repetition, thus, contains a GeO$_{2}$ sublayer and a SiO$_{2}$ sublayer. In the following experiments, the number of cycles in each sublayer is subsequently halved, while the number of repetitions is doubled in order to keep the total number of pulses constant. The XRR patterns of these films, their fits and an illustration of the models used in the fits, are plotted in Figure \ref{fig:SuperlatticeXRR} (the actual parameters of the model can be found in Table 1 of the Supplementary Material). By differentiating the experimental data (which has been obtained with a step size of 0.01º in 2$\theta$) at low angles and smoothing it with a Savitzky-Golay filter, we determined the critical angle for all the films to be 0.235º $\pm$ 0.005º, independent of the number of repetitions (or the periodicity of the multilayer), close to the bulk value for SiO$_{2}$ (0.234º for a density of 2.65g/cm$^3$).

The XRR patterns show clear thickness oscillations in all cases, indicating the good homogeneity of the films and the quality of the top and bottom interfaces. In addition, the patterns corresponding to the films containing from 2 up to 15 repetitions, all display superlattice reflections, attesting for the presence of chemical contrast between the SiO$_2$ and GeO$_2$ sublayers. The XRR pattern of the film with 15 repetitions shows its superlattice signature peak at about 7.1º, corresponding to a period of 12 Å, which is approximately the size of two unit cells in SiO$_2$ and GeO$_2$ crystalline polymorphs. In the case of the 15 repetitions sample, despite the clear superlattice peak, the model is not able to provide a reliable value for the thickness of the individual SiO$_{2}$ or GeO$_{2}$ sublayers. This can be understood looking at the roughness values (see Table 1 in Supplementary Material), which are of the order of the estimated sublayer thickness (6 Å). For the 30 repetitions film, only 5 pulses were provided, alternately, for each SiO$_{2}$ and GeO$_{2}$ sublayer, until completing the total of 300 pulses. Therefore, the superlattice periodicity is expected to be half of the value of the periodicity displayed by the 15 repetitions films, 6 Å. This value is similar to the roughness values obtained with the reflectivity fit and, thus, no chemical contrast is expected. Indeed, in this case, the superlattice model also gives unreliable results but, unlike in the case of the 15 repetition film, the 30 repetition XRR pattern can be modelled by a uniform layer (see Table 1 in Supplementary Material), consistent with the absence of superlattice peaks up to an angle of 14º.  

From this series of experiments it becomes clear that electron density, or composition, contrast between the SiO$_{2}$ and the GeO$_{2}$ sublayers is present down to the atomic level (close to a unit cell of their stable polymorphs). On the one hand, these results attest the excellent capabilities of ALD in general, allowing the growth of heterostructures with atomic-scale thickness control. On the other hand, it nicely illustrates a potential pitfall of the layered approach to compositional tuning in ALD, in that extremely short supercycles are needed in order to achieve a uniform composition rather than a superlattice.

As a final note, it must be pointed out that, as directly visible in Figure \ref{fig:SuperlatticeXRR}, the SiO$_{2}$:GeO$_{2}$ thickness ratio, and therefore the atomic ratio, changes from one film to the next, even though the pulse ratio for all of them is maintained as a 1:1. This dependence of the composition on the precise pulsing sequence, and not only on the pulse ratio, has been previously observed \cite{Longo2013}.

To conclude, we have successfully optimized the ALD growth of GeO$_{2}$ thin films from TDMAGe and O$_{3}$ precursors. In order to successfully achieve SiO$_{2}$/GeO$_{2}$ films by sequential layered growth, a compromise has been found between the optimal growth parameters and those parameters that will allow us to synthesize a mixed oxide film in a reasonable amount of time. Armed with this knowledge, we have set out on the synthesis of a series of increasingly intermixed SiO$_{2}$/GeO$_{2}$ thin films, showing that ALD indeed is able to achieve atomic level accuracy for these compounds as well. 

See the supplementary material for further details on the experimental methods, for the AFM images of the films, and for the details and fitting parameters giving rise to the XRR simulation curves in Figure 3. 

The data that support the findings of this study are available from the corresponding author upon reasonable request.

\section{\label{sec:Acknowledgements}Acknowledgements:\protect\\}

The authors are grateful to Picosun{\textregistered} for their optimization report related to the growth of SiO$_2$, to Adrian Carretero-Genevrier and Václav Ocelík for useful discussions and to Ir. Jacob Baas and the Zernike NanoLab Groningen for the technical support. The authors also acknowledge financial support from NWO’s TOP-PUNT grant 718.016002.

\section*{\label{sec:References}References:\protect\\}
\bibliography{library}

\end{document}